\documentclass[aps,prl,twocolumn,superscriptaddress,longbibliography]{revtex4-1}

\usepackage{pdfpages}
\usepackage{times}
\usepackage{amsmath,amssymb}
\usepackage{graphicx}
\usepackage{hyperref}
\usepackage{xcolor}

\makeatletter
\AtBeginDocument{\let\LS@rot\@undefined}
\makeatother

\begin{document}
	
\title{Information Scrambling in Quantum Neural Networks}

\author{Huitao Shen}
\affiliation{Department of Physics, Massachusetts Institute of Technology, Cambridge, Massachusetts 02139, USA}

\author{Pengfei Zhang}
\affiliation{Institute for Advanced Study, Tsinghua University, Beijing, 100084, China}
\affiliation{Institute for Quantum Information and Matter, California Institute of Technology, Pasadena, California 91125, USA}
\affiliation{Walter Burke Institute for Theoretical Physics, California Institute of Technology, Pasadena, California 91125, USA}

\author{Yi-Zhuang You}
\affiliation{Department of Physics, University of California, San Diego, California 92093, USA}

\author{Hui Zhai}
\email{hzhai@tsinghua.edu.cn}
\affiliation{Institute for Advanced Study, Tsinghua University, Beijing, 100084, China}

\begin{abstract}
The quantum neural network is one of the promising applications for near-term noisy intermediate-scale quantum computers. A quantum neural network distills the information from the input wavefunction into the output qubits. In this Letter, we show that this process can also be viewed from the opposite direction: the quantum information in the output qubits is scrambled into the input. This observation motivates us to use the tripartite information, a quantity recently developed to characterize information scrambling, to diagnose the training dynamics of quantum neural networks. 
We empirically find strong correlation between the dynamical behavior of the tripartite information and the loss function in the training process, from which we identify that the training process has two stages for randomly initialized networks. In the early stage, the network performance improves rapidly and the tripartite information increases linearly with a universal slope, meaning that the neural network becomes less scrambled than the random unitary. In the latter stage, the network performance improves slowly while the tripartite information decreases. We present evidences that the network constructs local correlations in the early stage and learns large-scale structures in the latter stage. We believe this two-stage training dynamics is universal and is applicable to a wide range of problems. Our work builds bridges between two research subjects of quantum neural networks and information scrambling, which opens up a new perspective to understand quantum neural networks. 
\end{abstract}
	
\maketitle

The neural network (NN) lies at the heart of the recent blossom of deep learning~\cite{goodfellow2016deep}. The NN distills information from the input, usually represented by a high-dimensional vector, and encodes it into a lower-dimensional output vector. Recently, quantum generalizations of NNs have been proposed and actively studied~\cite{PhysRevX.7.041052,Torrontegui_2019,Benedetti2019,Farhi2018,McClean2018,PhysRevA.98.032309,Huggins_2019,Schuld2018,Grant2018,PhysRevA.98.062324,Verdon2018,PhysRevA.99.052306,Du2018,Beer2019,PhysRevB.100.094434}. In a quantum NN, both the input and the output are quantum wavefunctions. The classical mapping is replaced by a quantum channel composed of unitary evolutions and measurements~\cite{Biamonte2017}.
The quantum NN is considered as one of the promising applications for near-term noisy intermediate-scale quantum devices~\cite{Preskill2018quantumcomputingin}. Moreover, it has been suggested that the quantum NN has more expressive power than its classical counterpart~\cite{Du2018}. 

\begin{figure}[tbp]
	\includegraphics[width=.85\columnwidth]{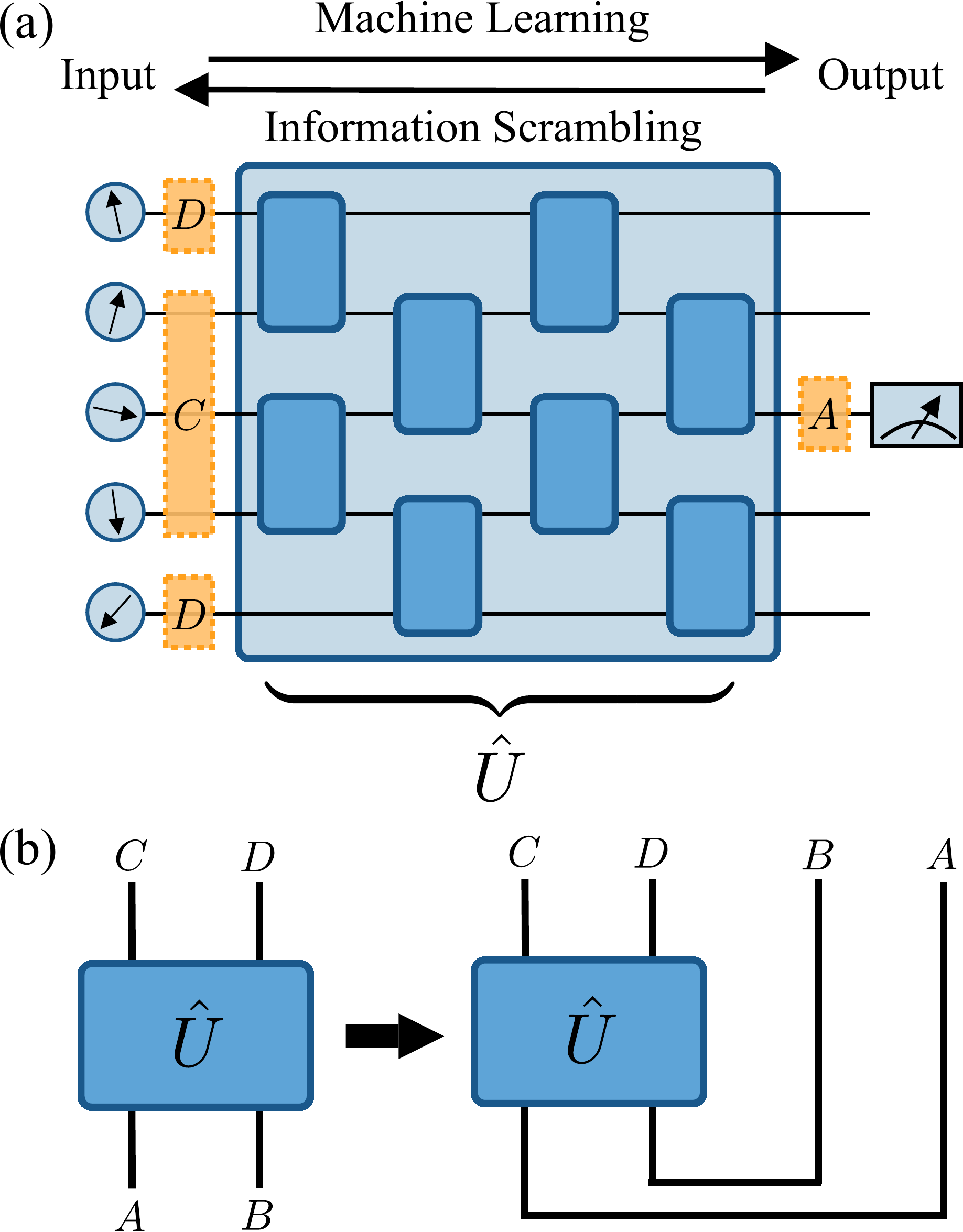}
	\caption{(a) Schematic of a quantum circuit with brick-wall geometry. Here the network has $n=5$ qubits and depth $l=4$. All two-qubit gates form a giant unitary transformation $\hat{U}$ that distills the information from the input qubits and encode it into one output qubit. The inverse process is that the information of one output qubit is scrambled into input qubits by $\hat{U}^\dag$. $A$ is the output subsystem, $C$ and $D$ are input subsystems in the definition of the tripartite information.  (b) Illustration for the operator-state mapping in the definition of tripartite information. Each leg may represent multiple qubits. } 
	\label{scheme}
\end{figure}

Similar to a classical NN, quantum information in the input wavefunction is distilled and encoded into the output in a quantum NN. This process is illustrated by the forward arrow in Fig.~\ref{scheme}(a). Intriguingly, for a quantum NN, this process can also be viewed from the opposite direction. By deferring measurements until the end of the quantum channel~\cite{nielsen2010quantum}, the information encoded in output qubits just before the measurement is spread into the entire system by unitary transformations, as illustrated by the backward arrow in Fig.~\ref{scheme}(a). Such processes that the information is scrambled from a small subsystem to a large one are known as the information scrambling.
The subject of information scrambling is well-studied in contexts such as thermalization, chaos and information dynamics in quantum many-body systems, and even black-hole physics~\cite{Altman2018,Qi2018,Swingle2018,Larkin1969,Kitaev2014,Shenker2014,Maldacena2016,FAN2017707}.

Quantum NNs and quantum information scrambling so far are two separated research topics. The purpose of this Letter is to bridge the gap and make their connection: In a quantum NN, information encoding and the information scrambling are the same process viewed from opposite directions. 

There have been information-theoretic studies of classical NNs~\cite{Schwartz2017,Saxe2018,goldfeld19a,Shen2019}. 
However, in classical NNs, the mapping at every layer is usually not invertible and the information is generally not preserved. Due to the information loss during the process, the mutual information always decreases with the network depth. 
In contrast, the unitarity of quantum evolutions preserves the information perfectly. The mutual information between the input and the output of any unitary transformation is always maximal. In order to have nontrivial diagnosis in quantum NNs, the key is to consider the mutual information between \textit{subsystems} of the input and the output. This naturally leads to the tripartite information---a quantity that characterizes the information scrambling~\cite{PhysRevLett.96.110404,Hosur2016}.

Here we study the training dynamics of quantum NNs using the tripartite information. We simultaneously monitor both the network performance and the tripartite information during training and observe empirical relations between them. Based on the behavior of these two quantities, the training process can be decomposed into two stages which we call the ``local construction stage'' and the ``global relaxation stage''. In the following, we present a detailed analysis of the training dynamics and provide evidence to support our claim.

\textit{Tripartite Information of Quantum Neural Networks.} 
Consider a unitary operator $\hat{U}$ in the $n$-qubit Hilbert space $\hat{U}=\sum_{i,j=1}^{2^n} U_{ij}|i\rangle \langle j|$, where $\{|i\rangle,i=1,\ldots,2^n\}$ denotes a complete set of bases in the Hilbert space. It can be regarded as a tensor with $n$ input and $n$ output legs. As illustrated in Fig.~\ref{scheme}(b), we divide the output legs (qubits) to two non-overlapping subsytems $A$ and $B$ and similarly divide the input legs (qubits) to $C$ and $D$.

The operator can be mapped to a state in the $2n$-qubit Hilbert space as $|U\rangle =\sum_{i,j=1}^{2^n} U_{ij}/\sqrt{2^n}|i\rangle |j\rangle$. Since $|U\rangle$ is a pure state, the entanglement entropy of its subsystem is well-defined, e.g. $S(A)\equiv -\mathrm{tr}(\rho_A\log_2 \rho_A)$ with $\rho_A\equiv \mathrm{tr}_{B,C,D}(|U\rangle\langle U|)$ being the reduced density matrix of subsystem $A$. The mutual information between the output subsystem $A$ and the input subsystem $C$ is $I(A,C)\equiv S(A)+S(C)-S(A\cup C)$. Similar definition can be made for $I(A,D)$ and $I(A,C\cup D)$. The tripartite information of the unitary $\hat{U}$ is defined as ~\cite{PhysRevLett.96.110404,Hosur2016}
\begin{align}
I_3(A,C,D)\equiv I(A,C)+I(A,D)-I(A,C\cup D),
\end{align}
Because $C\cup D$ are all input qubits, it can be proved that $I(A,C\cup D)=2|A|$, where $|A|$ is the number of qubits in subsystem $A$. Therefore, it is crucial to consider the mutual information between subsystems of both input and output qubits. 

The strong subadditivity of the entanglement entropy leads to $I_3(A,C,D)\leq 0$ for a unitary gate. The absolute value of the tripartite information $I_3(A,C,D)$ measures how much information of the subsystem $A$ is shared by $C$ and $D$ simultaneously after the unitary transformation, thus quantifies how scrambled a unitary is. For example, for an identity unitary transformation $U_{ij}=\delta_{ij}$, if $A$ is entirely contained in $C$ or $D$, it is straightforward to show that $I_3(A,C,D)=0$. As an opposite limit, for uniform Haar random unitary, local measurements cannot extract any information. It follows on average $I(A,C)$ and $I(A,D)$ are exponentially small and therefore $I_3(A,C,D)=-2|A|$, which is the minimal value for $I_3$~\cite{Hosur2016}. 

Having introduced the tripartite information for a general unitary transformation, we now turn to tripartite information of a quantum NN. Here we only consider parameterized quantum circuits with brick-wall geometry. As shown in Fig.~\ref{scheme}(a), each brick represents an independent two-qubit unitary gate in the $\mathrm{SU}(4)$ group, and is parameterized using its 15 Euler angles~\cite{Dita_2003}. During training, these parameters are optimized with classical optimization algorithms. All these two-qubit gates form a quantum circuit represented by a giant unitary transformation $\hat{U}$. 

The datasets to be studied in this work have several important features. First, the input wavefunctions all have time reversal symmetry, and consequently can be represented as real vectors. Therefore we restrict two-qubit gates to $\mathrm{SO}(4)$ with 6 Euler angles each. Second, the output target is either a real number within $[-1,1]$ or a binary label within $\{0,1\}$, only one readout qubit is needed at the end of the quantum circuit. For simplicity, we always let $n$ be odd and fix the readout qubit to be the qubit at the center, i.e. the $(n+1)/2$-th qubit. 

To define tripartite information, we always fix the output subsystem $A$ to be the central readout qubit. To respect the symmetry that $A$ is located at the center, we always choose $C$ to be the central $|C|$ input qubits in the circuit, and $D$ to be the remaining input qubits. Note that under this definition, $D$ in general contains two disconnected regions. The tripartite information $I_3(A,C,D)$ characterizes how much information of the output qubit is scrambled on the input side between the central region $C$ and the outer region $D$.

\textit{Magnetization Learning.} The first task is to supervisedly learn the average magnetization of a many-body wavefunction of $n$ half spins. The dataset consists of $N$ input-target pairs $\{(|G^\alpha\rangle,M_z^\alpha),\alpha=1,\ldots,N\}$, where the input wavefunction $|G^\alpha\rangle$ is the ground state wavefunction of the parent Hamiltonian with random long-ranged spin-spin interactions:
\begin{align}
\hat{H}=\sum_{i,j=1}^n(J_{ij}\sigma_i^z\sigma_j^z+K_{ij}\sigma_i^x\sigma_j^x)+\sum_{i=1}^n(g_i\sigma_i^x+h\sigma^z_i),
\label{H1}
\end{align}
where $\sigma^\mu_i$ represents the $\mu$-th Pauli matrix on the $i$-th qubit, $\mu=x,y,z$ and $i=1,\ldots,n$. $J_{ij}$, $K_{ij}$, $g_i$ and $h$ are all random numbers. The target is the average magnetization computed as $M_z^\alpha\equiv \langle G^\alpha|\hat{M}_z|G^\alpha\rangle$, where the magnetization operator is $\hat{M}_z\equiv \sum_{i=1}^n \sigma_i^z/n $. In sampling the random Hamiltonian, we ensure $J_{ij}\leq 0$ such that the ground state wavefunctions are either ``ferromagnetic'' or ``paramagnetic'' measured under $\hat{M}_z$. $h$ is a small pinning field randomly drawn from a distribution with zero mean, which is used to trigger the spontaneous $Z_2$ symmetry breaking in the ferromagnetic phase.

The quantum NN takes the input wavefunction $|G^\alpha\rangle$ and applies the unitary transformation $\hat{U}$ on it. The magnetization is readout by measuring $\sigma^x$ of the central qubit. We choose to measure $\sigma^x$ instead of $\sigma^z$ because the quantum NN may learn some shortcut that is unable to generalize if the measurement and the target physical observable are under the same basis. This is essentially a regression task and the loss function to be minimized is the absolute error of the magnetization:
\begin{align}
\mathcal{L}=\frac{1}{N}\sum_{\alpha=1}^N \left|\left\langle G^\alpha\right|\hat{U}^\dagger\sigma_{(n+1)/2}^x\hat{U}\left|G^\alpha\right\rangle-M_z^\alpha\right|.
\end{align}

We simulate the above hybrid quantum-classical quantum NN training algorithm. The distributions of random parameters in the Hamiltonian Eq.~\eqref{H1} are chosen such that $M_z^\alpha$ in the dataset roughly distributes uniformly within $[-1,1]$. 
All two-qubit unitaries in the quantum NN are initialized randomly. The parameters are optimized with the AMSGrad gradient descent algorithm~\cite{Sashank2018}. The gradients can be computed directly thanks to the linearity of the quantum channel and are measurable in a realistic quantum NN~\cite{SM,PhysRevA.98.032309,Schuld2018}.  

\begin{figure}[tbp]
	\includegraphics[width=.9\columnwidth]{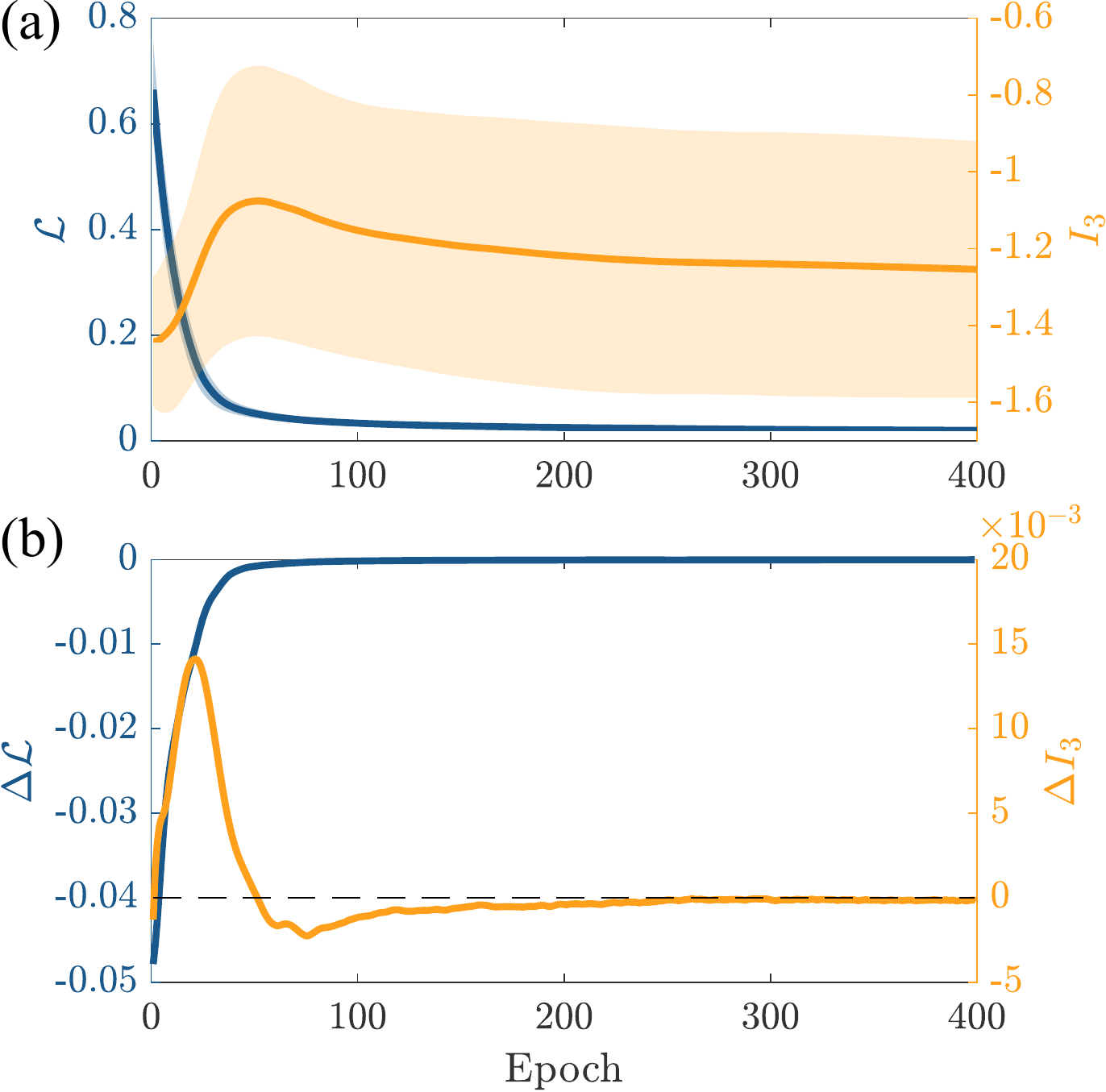}
	\caption{(a) Training loss $\mathcal{L}$ and tripartite information $I_3(A,C,D)$ as functions of the training epoch. The shaded area represents one standard deviation. (b) Finite difference of training loss $\Delta \mathcal{L}$ and tripartite information $\Delta I_3$ as functions of the training epoch. The dotted vertical line indicates the boundary between two training stages, which is determined as the maximum of the averaged $I_3$ given by $\Delta I_3=0$. All results are averaged over 20 different random initializations. The network has $n=9$ qubits and depth $l=6$. The training and validation dataset contains $N=2500$ and $500$ wavefunction--magnetization pairs respectively, sampled from random Hamiltonian ensemble, where random parameters are distributed uniformly within $J_{ij}/J\in [-1,0]$, $K_{ij}/J\in [-1,1]$, $g_i/J\in[-6,6]$ and $h/J\in[-0.04,0.04]$ . $J$ is the energy unit. The learning rate is $\lambda =10^{-2}$. Here and in the rest of the paper the input subsystem size $|C|=5$. }
	\label{time_scales}
\end{figure}

\textit{Two-stage Training.} In Fig.~\ref{time_scales}(a), we show the training loss and the tripartite information, both averaged over different initilizations, as functions of the training epoch. Averaging over different initializations reduces the volatility within a single training instance and makes the correlation between the two quantities clearer.
At the early stage of the training, the rapid improvement of the quantum NN performance, characterized by a fast decrease of the training loss, is accompanied by an almost linear increase of the tripartite information. In other words, the quantum NN becomes less scrambled compared with the initial random unitary. This training stage terminates when the tripartite information reaches its local maximum. In the next stage, the tripartite information decreases again, meaning that the network scrambles information faster. The network performance also improves, but with a much slower rate compared with that in the first stage. In Fig.~\ref{time_scales}(b), we plot the finite difference of the two metrics $\Delta\mathcal{L}$ and $\Delta I_3$ together, and use a dashed line to indicate the maximum of $I_3$ given by $\Delta I_3=0$. One can see clearly that $\Delta\mathcal{L}$ also drops to negligible small values around the dashed line, meaning a much slower decreasing rate of $\mathcal{L}$ in the later stage.

We call the training stage before $I_3$ reaching the maximum the ``local construction stage'', and the latter stage where $I_3$ decreases as the ``global relaxation stage''. The reason for the names will be clear after we study the training dynamics in detail below. The empirical observation that quantum NN performance and the information scrambling is closely correlated is the main finding of this work. This correlation has been observed in all our numerical simulations with different network initializations, training algorithms, system sizes and network depths~\footnote{For network initializations, we require initial unitaries to be scrambled enough such that initial $I_3(A,C,D)\lesssim -1$ ($-1$ is about is half of the negative-most value). For training algorithms, we require these algorithms to be gradient-based. For network depths, we require the networks to be not too shallow. }. 
We also train quantum NNs for learning the staggered magnetization from the ground state of random antiferromagnetic and even frustrated Hamiltonians, and the winding number of a product quantum state. 
Despite the very different nature of these tasks, the empirical correlation between the NN performance and the tripartite information still holds. All details are presented in \cite{SM}.

\begin{figure}[tbp]
 	\center
 	\includegraphics[width=.9\columnwidth]{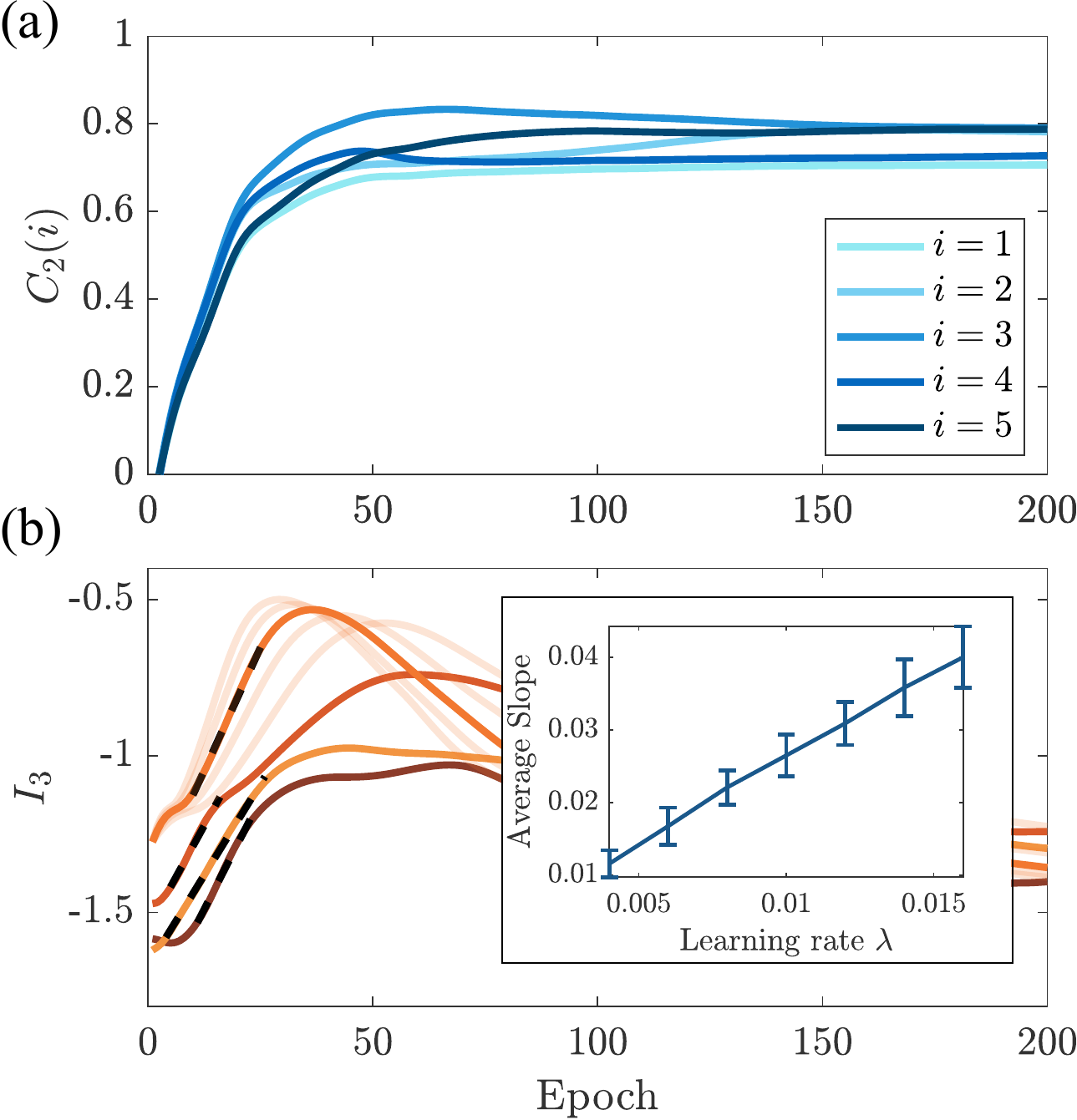}
 	\caption{(a) Two-point correlation function $C_2(i)$ as a function of the training epoch and the input qubit $i$ for a typical initialization. (b) Tripartite information $I_3(A,C,D)$ as a function of the training epoch for different initializations and learning rates. All solid lines are trained under learning rate $\lambda=10^{-2}$. The transparent orange lines are trained with the same initialization as the solid orange line, but with learning rates  $\lambda=6,8,12,14\times 10^{-3}$. 
 	The average slope for the four initializations shown here is plotted in the inset, as a function of the learning rates. The error bars represent the standard deviation of fitted slopes for fixed learning rate but different initializations. } 
 	\label{local_time}
 \end{figure}

\textit{Local Construction Stage.} We claim that during the first stage when the tripartite information linearly increases, the quantum NN learns local features of the input wavefunction. For the magnetization learning task, because of the existence of ferromagnetic domain, there is some probability that any single spin is aligned relatively well with remaining spins in the system. Simply outputting any single-spin magnetization of the input wave function is actually a reasonable guess, so that the training loss can decrease rapidly. For such networks where only local features are extracted, information does not need to be scrambled into the whole system. Therefore, the tripartite information increases during this stage. 

To support the above claim, we compute two-point correlations between input qubits and the readout qubit:
\begin{align}
C_2(i)\equiv \frac{1}{N}\sum_{\alpha=1}^N \left\langle G^\alpha\right|\sigma^z_i \hat{U}^\dagger\sigma_{(n+1)/2}^x\hat{U}\left|G^\alpha\right\rangle.
\end{align}
If one views $\hat{U}$ as a time evolution operator, then $C_2(i)$ is simply a two-point function between two different places and two different times. In Fig.~\ref{local_time}(a), we plot $C_2$ as a function of different input qubits and training epochs in the early training stage. As can be seen, they increase rapidly and then saturate to large values. The increasing correlation indicates that the quantum NN is establishing the correspondence between local input features and the output qubit. During this stage, the tripartite information also increases, and the two-point correlation function saturates when the tripartite information reaches the maximum. All these observations are consistent with our claim that during the first local construction stage, local features are extracted from the input. 

Before concluding this section, we point out another interesting observation that the linear increasing slope of the tripartite information is nearly a constant that is independent of the initialization, shown in Fig.~\ref{local_time}(b). Of course, this slope depends on the learning rate of the gradient descent algorithm. As shown in the inset, the $I_3$-independent slope scales linearly with the learning rate.

\begin{figure}[tbp]
	\center
	\includegraphics[width=.9\columnwidth]{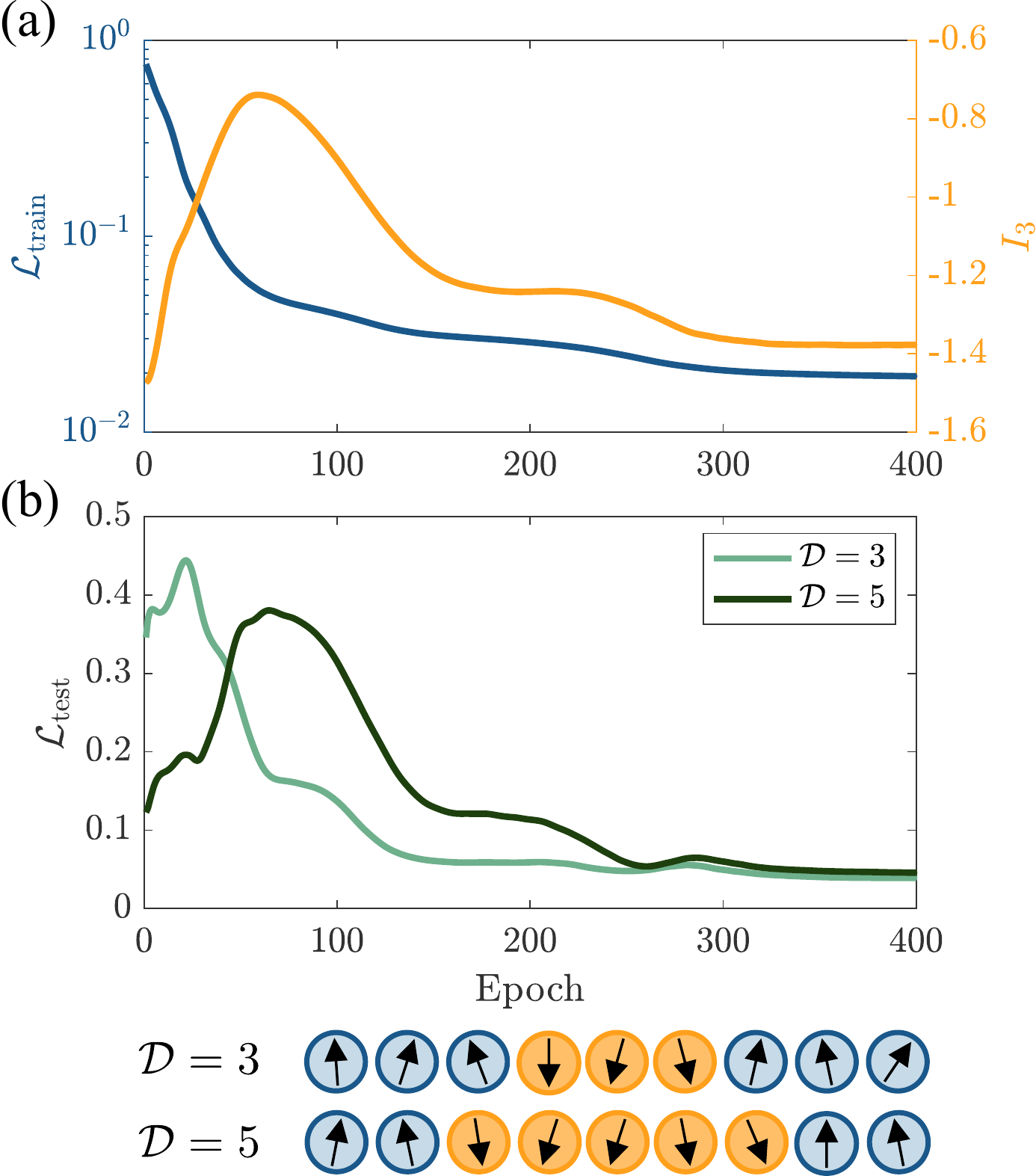}
	\caption{(a) Training loss and tripartite information as functions of the training epoch for a typical initialization. (b) Loss functions on the artificial test dataset with ``ferromagnetic domain'' of size $\mathcal{D}=3$ and $5$ for the same training instance as (a). } 
	\label{global_time}
\end{figure}

\textit{Global Relaxation Stage.} We now turn to the second stage where the tripartite information decreases and the training loss decreases with a much slower rate. We claim that during this stage, the quantum NN learns global features of the wavefunction. 
To provide evidence for this claim, we test the quantum NN in an artificial test dataset $\{(|\psi_{\mathcal{D}}^\alpha\rangle,M_z^\alpha),\alpha=1,\ldots,N_{\mathcal{D}}\}$, constructed according to the following process. First, we sample ground states $|G^\alpha\rangle$ from the random Hamiltonian of Eq.~\eqref{H1}. Next, we apply the following unitary transformation to flip a region of spins:
\begin{equation}
|\psi_\mathcal{D}^\alpha\rangle=\prod_{\frac{n-\mathcal{D}+1}{2} \leq i\leq \frac{n+\mathcal{D}}{2}}\sigma^x_i|G^\alpha\rangle.
\end{equation}
For ``paramagnetic'' wavefunctions $|G^\alpha\rangle$, this transformation leaves these wavefunctions ``paramagnetic''. However, for ``ferromagnetic'' wavefunctions $|G^\alpha\rangle$, the transformation creates a ferromagnetic domain wall of size $\mathcal{D}$, as sketched in Fig.~\ref{global_time} . In order to accurately compute the magnetization of such wavefunctions, the quantum NN must be able to learn structures larger than the domain wall size $\mathcal{D}$. In \cite{SM}, we present an argument on why in this task, long string operators should exist in $\hat{U}^\dagger\sigma_{(n+1)/2}^x\hat{U}$ when it is expanded under the basis of product of local Pauli matrices.

In Fig.~\ref{global_time}(b), we show losses on test datasets with $\mathcal{D}=3$ and $5$, as functions of the training epoch. 
In the later stage of training, although the training loss decreases slowly, the tripartite information can decrease rather drastically, accompanied by a rapid decreasing of losses on both test datasets. Moreover, the larger the domain wall size is, the later the test loss begins to decrease. This means that the information scrambling is associated with the performance improvement on wavefunctions with large domain structures. This naturally explains why the unitary has to become more scrambled. Since such data are rare in the training dataset, it also explains why the training loss improvement is slow. 
Finally, we note that in Fig.~\ref{time_scales}, the standard deviation of $I_3$ is quite large in the later stage. This is consistent with the chaotic nature of the information scrambling, as it is now known that the quantum many-body chaos and the information scrambling are two closely related concepts. 

\textit{Discussion and Outlook.} In summary, we apply a metric of quantum information scrambling---the tripartite information---to diagnose the training process of quantum NNs. We find strong correlation between this metric and the loss function, and identify a two-stage training dynamics of quantum NN. We show that the quantum NN establishes local correlations in the early stage and builds up global structures in the later stage. Such two-stage dynamics is reminiscent of physical processes such as annealing of ferromagnetism, and the operator growth in many-body quantum chaos. We believe this two-stage dynamics is universal for a wide range of quantum machine learning problems. We also believe that the profound connection between the information scrambling and the quantum NN could find broader applications in quantum machine learning, such as revealing the underlying mechanism of quantum machine learning and guiding the quantum NN architecture design.


\textit{Acknowledgment.} We thank Yingfei Gu for discussions and an anonymous referee for the suggestion to average results from different initializations. HS thanks IASTU for hosting his visit to Beijing, where key parts of this work were done. HS thanks Guangyu Du for suggestions on the data presentation. PZ acknowledges support from the Walter Burke Institute for Theoretical Physics at Caltech. This work is supported by Beijing Outstanding Young Scientist Program (HZ), MOST under Grant No.~2016YFA0301600 (HZ) and NSFC Grant No.~11734010 (HZ).

\bibliography{Reference.bib}

\widetext
\clearpage


\section*{Supplementary material}

# Supplemental Material for "Information Scrambling in Quantum Neural Networks"

Huitao Shen,[1] Pengfei Zhang,[2] Yi-Zhuang You,[3] and Hui Zhai[2, *]

[1]*Department of Physics, Massachusetts Institute of Technology, Cambridge, Massachusetts 02139, USA*
[2]*Institute for Advanced Study, Tsinghua University, Beijing, 100084, China*
[3]*Department of Physics, University of California, San Diego, CA 92093, USA*

In this supplemental material, we present more results of magnetization learning, staggered magnetization learning, and winding number learning, along with details of gradient calculation and measurement.

## I. MAGNETIZATION LEARNING

In this section, we provide more details of magnetization learning and present an argument on why in magnetization learning, long string operators should exist in $\hat{U}^\dagger \sigma^x_{(n+1)/2} \hat{U}$ when it is expanded under the basis of product of local Pauli matrices.

### A. Learning Task Details

Figure 1 shows the distribution of magnetization $M_z^\alpha$ in the training and validation datasets. The magnetization distributions within the training and validation set are similar. There are roughly equal number of wavefuntions that are "ferromagnetic" ($|M_z^\alpha| \geq 0.5$) or "paramagnetic" ($|M_z^\alpha| < 0.5$).

For the AMSGrad algorithm [1], momentum parameters are always $\beta_1 = 0.9$ and $\beta_2 = 0.999$ throughout this work. Becasue the training set is not very big, we use gradient descent instead of stochastic or mini-batch gradient descent. In other words, each epoch involves only one gradient descent step.

We confirm that validation losses also decrease monotonically when the training proceeds (not shown here), indicating that the network can learn to compute the magnetization reasonably well without overfitting.

In Fig. 2, we show the training loss and the tripartite information as functions of the training epoch. We plot both the averaged values over 20 different random initializations and two typical initializations. Both the averaged value and the two training instances show two-stage training dynamics. In particular, Fig. 3(a) and Fig. 4 in the main text use the same initialization as Initialization 2 here.

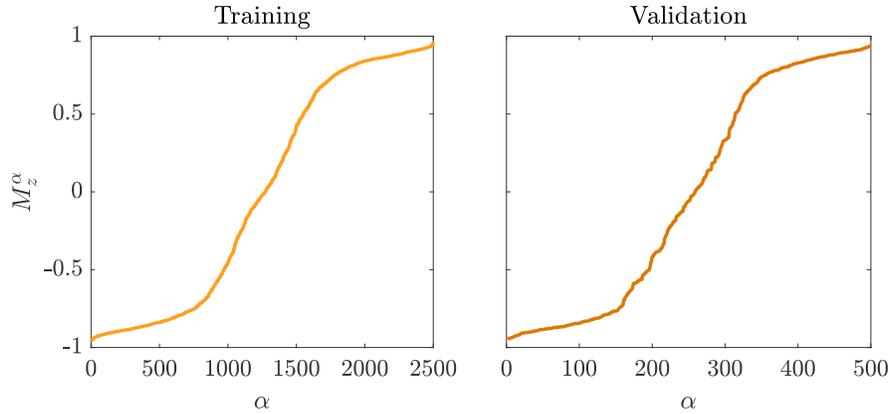

FIG. 1. Distribution of magnetization $M_z^\alpha$ in the training and validation sets. The training and validation dataset contains $N = 2500$ and $500$ wavefunction–magnetization pairs respectively, sampled from random Hamiltonian ensemble of system size $n = 9$, where random parameters are distributed uniformly within $J_{ij}/J \in [-1, 0]$, $K_{ij}/J \in [-1, 1]$, $g_i/J \in [-6, 6]$ and $h/J \in [-0.04, 0.04]$ . $J$ is the energy unit.

---

* hzhai@tsinghua.edu.cn

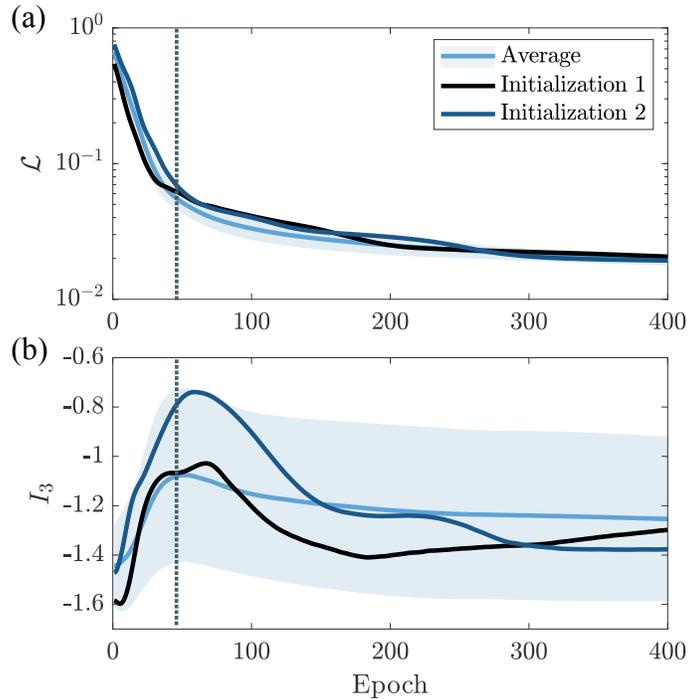

FIG. 2. **Magnetization learning.** (a) Training loss as functions of the training epoch. Different colors represent the average over 20 different random initializations or typical results from two training instances. The shaded area represents one standard deviation. The network has $n = 9$ qubits and depth $l = 6$. The learning rate is $\lambda = 10^{-2}$. (b) Tripartite information $I_3(A, C, D)$ as a function of the training epoch. Here the input subsystem size $|C| = 5$. The dotted vertical line indicates the boundary between two training stages, which is determined as the local maximum of the averaged $I_3$.

## B. Explicit Construction of Unitary that Learns Magnetization

Generally, it is impossible to find an unitary $\hat{U}$ such that

$$\hat{U}^\dagger \sigma^x_{(n+1)/2} \hat{U} = \hat{M}_z, \tag{1}$$

because the L.H.S. and the R.H.S. of the above equality have different eigenvalues. As a result, we can only expect the above equality to hold at the level of expectation

$$\langle \psi | \hat{U}^\dagger \sigma^x_{(n+1)/2} \hat{U} | \psi \rangle = \langle \psi | \hat{M}_z | \psi \rangle, \tag{2}$$

within a subset of states $\{|\psi\rangle\}$ that are of interest [1]. In the following, we present an explicit construction of $\hat{U}^\dagger \sigma^x_{(n+1)/2} \hat{U}$ for the magnetization learning problem when the subset of states are eigenstates of $\hat{M}_z \equiv \sum_{i=1}^{n} \sigma^z_i / n$. The purpose of this construction is to use an explicit example to demonstrate why it is usually necessary to have string operators in $\hat{U}^\dagger \sigma^x_{(n+1)/2} \hat{U}$ for a quantum neural netowrk (NN) that learns magnetization.

We first elaborate the rationale behind choosing eigenstates of $\hat{M}_z$. For magnetization learning, the dataset consists of ground states of Hamiltonian (Eq. (2) in the main text) with a small pinning field $h$ to trigger spontaneous $Z_2$ symmetry breaking in the finite-size numerical simulation. The spin-spin interaction is also chosen to be nonlocal to ensure that we have sufficient number of distinct states. To actually probe the physics of $Z_2$ symmetry breaking in one dimension, we should take the thermodynamics limit $n \to \infty$ while sending $h \to 0$ and fixing the spin-spin interaction range. It is well-known that in such systems, the ordered ferromagnetic ground state is gapped. Consequently, the quantum fluctuation of $\hat{M}_z$ is

$$\left\langle \left( \hat{M}_z - \left\langle \hat{M}_z \right\rangle \right)^2 \right\rangle = \sum_{i,j=1}^{n} \frac{1}{n^2} \left\langle \delta \sigma^z_i \delta \sigma^z_j \right\rangle = \sum_{i=1}^{n} \frac{1}{n} \left\langle \delta \sigma^z_i \delta \sigma^z_1 \right\rangle \sim \frac{1}{n}, \tag{3}$$

---

[1] Note that the subset is in general not a subspace as linear combinations in general break the equality Eq. (2).

because $\langle \delta \sigma_i^z \delta \sigma_1^z \rangle$ decays exponentially with $i$. Therefore, the fluctuation of $\hat{M}_z$ is suppressed, and ground states of our random Hamiltonian can be well approximated by eigenstates of $\hat{M}_z$ in the thermodynamic limit.

We are now ready to present our construction. Denote the eigenstates of $\hat{M}_z$ as $|m, i\rangle$ such that $\hat{M}_z |m, i\rangle = m |m, i\rangle$. Here $m \in [-1, 1]$ is the eigenvalue, which is also the average magnetization. $i = 1, \ldots, d_m$ represents the state in the degenerate eigenspace and $d_m$ is the degeneracy. The states are orthonormal $\langle m, i | m', i' \rangle = \delta_{mm'} \delta_{ii'}$ and complete $\sum_m d_m = 2^n$. Because of the spin-flip symmetry, $d_m = d_{-m}$. In general $d_m > 1$ unless $m = \pm 1$, where all spins are polarized to the same direction. For degenerate subspaces, note that the choice of $|m, i\rangle$ for fixed $m$ but different $i$ is not unique.

In the following, we construct matrix elements of $\hat{U}^\dagger \sigma_{(n+1)/2}^x \hat{U}$ under $|m, i\rangle$ basis such that

$$\langle m, i | \hat{U}^\dagger \sigma_{(n+1)/2}^x \hat{U} | m, i \rangle = m, \tag{4}$$

for all $m$ and $i$. Consider the two-dimensional subspace spanned by $|m, i\rangle$ and $|-m, i\rangle$ for all $m$ and $i$. Within this subspace, we set

$$\hat{U}^\dagger \sigma_{(n+1)/2}^x \hat{U} = \sin \theta \sigma^x + \cos \theta \sigma^z, \tag{5}$$

where $\theta = \arccos m$. It is straightforward to verify the constraint Eq. (4) is satisfied and the eigenvalues are $\pm 1$. Under this construction, half of $\hat{U}^\dagger \sigma_{(n+1)/2}^x \hat{U}$'s eigenvalues are $+1$ and half are $-1$. It is then not hard to see that there must exist some $\hat{U}$ such that $\hat{U}^\dagger \sigma_{(n+1)/2}^x \hat{U}$ has the matrix elements under $|m, i\rangle$ basis as constructed.

Although the above matrix is constructed explicitly on a particular choice of basis, it is straightforward to verify that the following basis-independent constraint holds

$$\langle m | \hat{U}^\dagger \sigma_{(n+1)/2}^x \hat{U} | m \rangle = m, \tag{6}$$

where $|m\rangle \equiv \sum_{i=1}^{d_m} c_i |m, i\rangle$ is any linear combination of eigenstates within the same degenerate eigenspace. $\sum_{i=1}^{d_m} |c_i|^2 = 1$.

Because the choice of basis within a degenerate subspace is not unique, our constructions above are not unique either. Nevertheless, generally $|m, i\rangle$ and $|-m, i\rangle$ are related to each other by a string of local Pauli matrices whose length is of order of system size $n$. A particular choice is that $|-m, i\rangle = \prod_{j=1}^n \sigma_i^x |m, i\rangle$ such that the two states are related by a global spin-flip operator, which is a string operator of length $n$. Because of Eq. (5), such string operator must exist in $\hat{U}^\dagger \sigma_{(n+1)/2}^x \hat{U}$.

## II. STAGGERED MAGNETIZATION LEARNING

In this section, we present results of staggered magnetization learning task, where empirical correlation between the NN performance and the tripartite information is also found.

*Dataset.* Similar to magnetization learning, the dataset consists of $N$ input-target pairs $\{(|G^\alpha\rangle, \overline{M}_z^\alpha), \alpha = 1, \ldots, N\}$, where the input wavefunction $|G^\alpha\rangle$ is the ground state wavefunction of the parent Hamiltonian with random long-ranged spin-spin interactions:

$$\hat{H} = \sum_{i=1}^{n-1} J_{i,i+1} \sigma_i^z \sigma_{i+1}^z + \sum_{i,j=1}^n K_{ij} \sigma_i^x \sigma_j^x + \sum_{i=1}^n (g_i \sigma_i^x + h \sigma_i^z), \tag{7}$$

where $J_{i,i+1}$, $K_{ij}$, $g_i$ and $h$ are all random numbers. Compared with Eq. (2) in the main text, here only nearest-neighbour $\sigma_i^z \sigma_{i+1}^z$ interactions are included to avoid frustration.

The target is the average staggered magnetization computed as $\overline{M}_z^\alpha \equiv \langle G^\alpha | \hat{\overline{M}}_z | G^\alpha \rangle$, where the staggered magnetization operator is $\hat{\overline{M}}_z \equiv \sum_{i=1}^n (-1)^n \sigma_i^z / n$. In sampling the random Hamiltonian, we ensure $J_{ij} \geq 0$ such that the ground state wavefunctions are either "antiferromagnetic" or "paramagnetic" measured under $\hat{\overline{M}}_z$. $h$ is a small pinning field randomly drawn from a distribution with zero mean, which is used to trigger the spontaneous $Z_2$ symmetry breaking in the antiferromagnetic phase.

*Task.* The quantum NN takes the input wavefunction $|G^\alpha\rangle$ and applies the unitary transformation $\hat{U}$ on it. The staggered magnetization is readout by measuring $\sigma^x$ of the central qubit. The loss function to be minimized is the absolute error of the staggered magnetization:

$$\mathcal{L} = \frac{1}{N} \sum_{\alpha=1}^N \left| \langle G^\alpha | \hat{U}^\dagger \sigma_{(n+1)/2}^x \hat{U} | G^\alpha \rangle - \overline{M}_z^\alpha \right|. \tag{8}$$

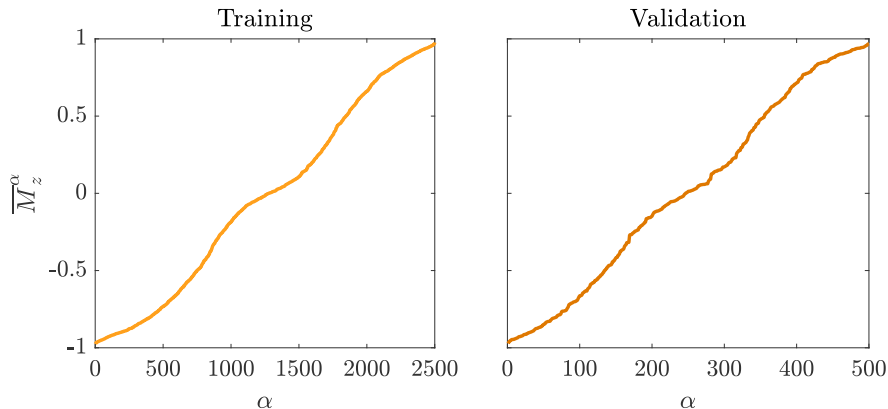

FIG. 3. Distribution of staggered magnetization $\overline{M}_z^\alpha$ in the training and validation sets. The training and validation dataset contains $N = 2500$ and 500 wavefunction–staggered magnetization pairs respectively, sampled from random Hamiltonian ensemble of system size $n = 9$, where random parameters are distributed uniformly within $J_{i,i+1}/J \in [0,8]$, $K_{ij}/J \in [-1,1]$, $g_i/J \in [-2,2]$ and $h/J \in [-0.04, 0.04]$ . $J$ is the energy unit.

In Fig. 3, we show the distribution of staggered magnetization $\overline{M}^\alpha$ in the training and validation datasets. The magnetization distributions within the training and validation set are similar. There are roughly equal number of wavefuntions that are "antiferromagnetic" ($|\overline{M}_z^\alpha| \geq 0.5$) or "paramagnetic" ($|\overline{M}_z^\alpha| < 0.5$).

*Results.* Figure 4 is the training loss and tripartite information during quantum NN training for the staggered magnetization learning task. We confirm the validation loss is similar to that in the training set.

The figure looks almost identical to Fig. 2, despite that we now have a different dataset. The two-stage training dynamics, i.e., an early stage with rapid decrease of loss and increase of tripartite information, followed by a later stage with slow decrease of both loss and tripartite information, can be clearly observed. One can also see the initial rapid linear growth of tripartite information in Fig. 4 with almost identical slopes for both two training instances and the averaged result.

Finally, we have also tried Hamiltonians similar to Eq. (7) but with longer interaction range such that the ground state is frustrated. The quantum NN shows similar performance and training dynamics.

## III. WINDING NUMBER LEARNING

In this section we present the results of winding number learning task, which again reinforces the generality of two-stage training dynamics of quantum NNs.

*Dataset.* The input data consist of $N$ product states of $n$ qubits, where each qubit represents a vector on the $xz$ plane of the Bloch sphere. The target is the winding number of these vectors by treating the $n$ qubits as vectors on an one-dimensional Brillouin zone [2]. Formally, the dataset consists of $N$ input-target pairs $\{(|H^\alpha\rangle, w^\alpha), \alpha = 1, \ldots, N\}$, where the input wave-function $|H^\alpha\rangle = \prod_{i=1}^n |\psi^\alpha(k_i)\rangle$, $k_i = 2\pi(i-1)/(n-1)$, and $\psi^\alpha(k)$ is the ground state of the following random two-band Hamiltonian in one-dimensional Brillouin zone $k \in [0, 2\pi)$ with chiral symmetry $\sigma_y H(k)\sigma_y = -H(k)$:

$$H(k) = h_x(k)\sigma^x + h_z(k)\sigma^z. \tag{9}$$

Here the coefficient $h_\mu(k)$, $\mu = x, z$ is represented in terms of Fourier components up to $p$-th harmonic:

$$h_\mu(k) = \sum_{n=0}^{p} \cos(nk)c_n^\mu + \sum_{n=1}^{p} \sin(nk)s_n^\mu, \tag{10}$$

where $c_n^\mu$ and $s_n^\mu$ are random numbers.

The learning target is the discrete version of winding number:

$$w^\alpha = \frac{1}{2\pi} \sum_{i=1}^{n} \text{Im} \ln \left[ e^{i(\phi^\alpha(k_i) - \phi^\alpha(k_{i+1}))} \right], \tag{11}$$

where $\phi^\alpha(k)$ is defined as the argument of the following complex number:

$$e^{i\phi^\alpha(k)} = \frac{h_z^\alpha(k) + ih_x^\alpha(k)}{\sqrt{h_z^\alpha(k)^2 + h_x^\alpha(k)^2}}. \tag{12}$$

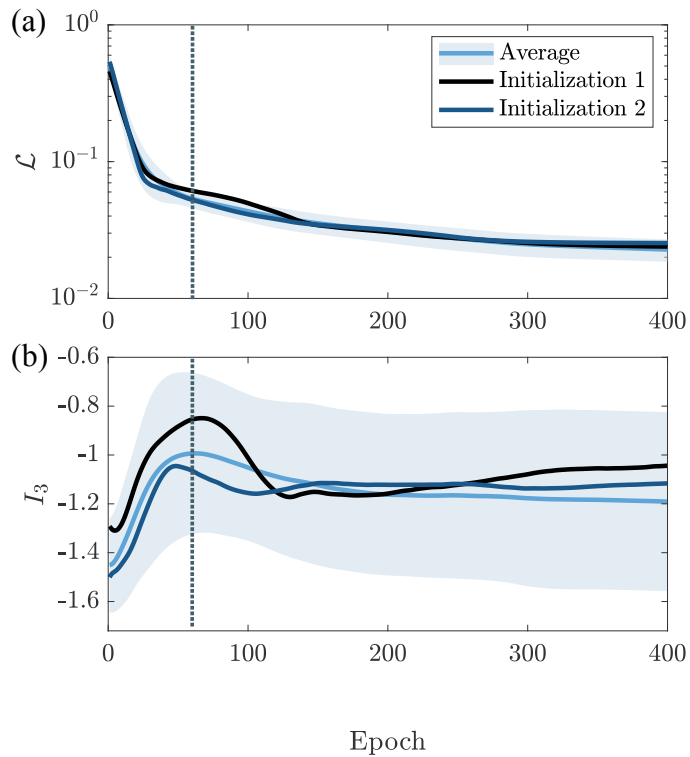

FIG. 4. Staggered magnetization learning. (a) Training loss as functions of the training epoch. Different colors represent the average over 20 different random initializations or typical results from two training instances. The shaded area represents one standard deviation. The network has $n = 9$ qubits and depth $l = 6$. The learning rate is $\lambda = 10^{-2}$. (b) Tripartite information $I_3(A, C, D)$ as a function of the training epoch. Here the input subsystem size $|C| = 5$. The dotted vertical line indicates the boundary between two training stages, which is determined as the local maximum of the averaged $I_3$.

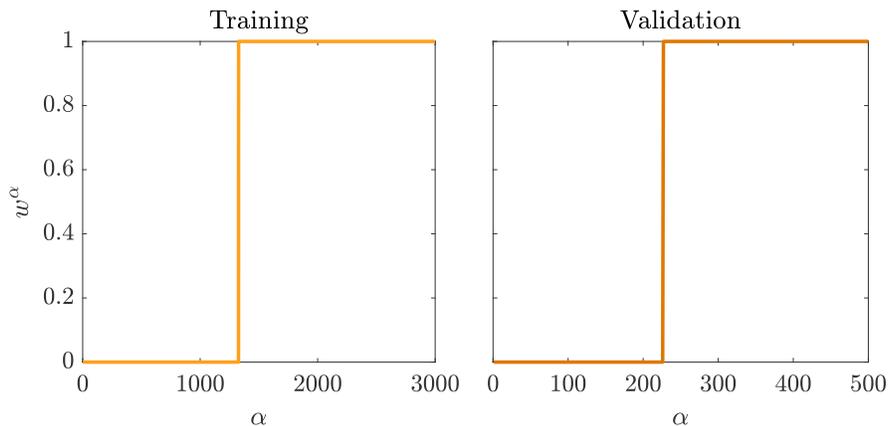

FIG. 5. Distribution of winding number $w^{\alpha}$ in the training and validation sets.

The branch cut for the logarithm in Eq. (11) is along the negative direction of the $x$ axis such that $\phi(k) - \phi(k') \in [-\pi, \pi)$.

*Task.* In the following, we set the harmonic cutoff $p = 1$. $c_n^{\mu}$ and $s_n^{\mu}$ are sampled from a uniform distribution between $[-1/3, 1/3]$ for $n = 0$ and $[-1, 1]$ for $n > 0$. We then post-select data with winding number $w = 0, 1$ and discard those with $w = -1$. In this way, the task becomes binary classification. The parameters are chosen such that there are roughly equal number of data with $w = 0$ and 1, as shown in Fig. 5.

The quantum NN takes the input wavefunction $|H^{\alpha}\rangle$ and applies the unitary transformation $\hat{U}$ on it. The probability that the

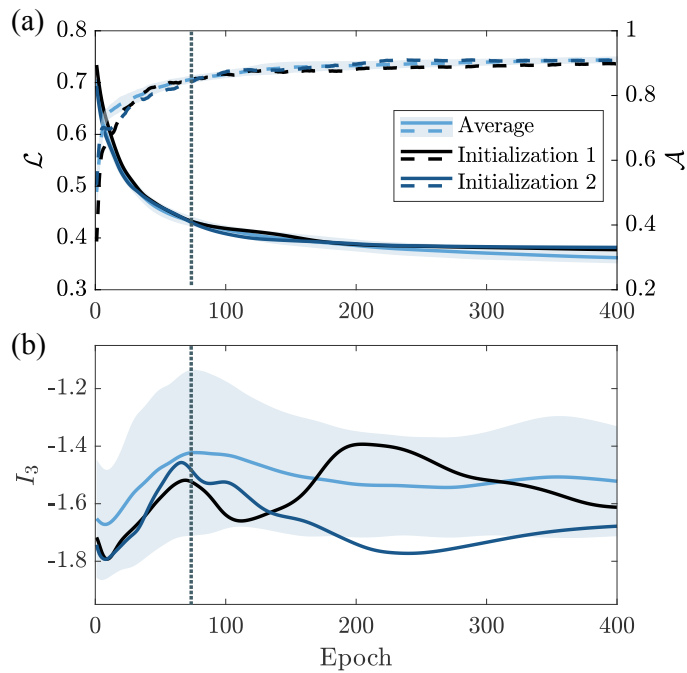

FIG. 6. Winding number learning. (a) Training loss (solid, left) and accuracy (dashed, right) as functions of the training epoch. Different colors represent the average over 20 different random initializations or typical results from two training instances. The shaded area represents one standard deviation. The network has $n = 9$ qubits and depth $l = 8$. The training and validation dataset contains $N = 3000$ and $500$ wavefunction-winding number pairs respectively, sampled from random wavefunctions defined in the main text. The learning rate is $\lambda = 10^{-2}$. (b) Tripartite information $I_3(A, C, D)$ as a function of the training epoch for different initializations. Here the input subsystem size $|C| = 5$. The dotted vertical line indicates the boundary between two training stages, which is determined as the local maximum of the averaged $I_3$.

$w^\alpha = 1$ is readout by measuring $\sigma^x$ of the central qubit:

$$p^\alpha = \frac{1 + \langle H^\alpha | \hat{U}^\dagger \sigma^x_{(n+1)/2} \hat{U} | H^\alpha \rangle}{2}. \tag{13}$$

Therefore, the loss function to be minimized is the negative binary cross-entropy:

$$\mathcal{L} = \frac{1}{N} \sum_{\alpha=1}^{N} \left[ -w^\alpha \ln p^\alpha - (1 - w^\alpha) \ln(1 - p^\alpha) \right]. \tag{14}$$

A more sensible metric is the prediction accuracy. Let the prediction of the winding number be $o^\alpha \equiv (1 + \mathrm{sgn}(p_a - 1/2))/2$. The prediction accuracy is then

$$\mathcal{A} \equiv 1 - \frac{1}{N} \sum_{\alpha=1}^{N} |o^\alpha - w^\alpha|. \tag{15}$$

*Results.* In Fig. 6, we present the training loss and accuracy for the winding number learning task, along with the tripartite information. We confirm the validation loss and accuracy is similar to that in the training set. The network depth $l$ is larger than that in the magnetization learning as we suspect the winding learning task is more difficult. However, using a shallower network will not affect the performance significantly. Because of the difficulty of this task, not all initializations can lead to high accuracies after 400 epochs. In computing the average, we post-select 20 different initializations with smallest training losses out of 50 initializations.

First, the quantum NN manages to learn distinguish wavefunctions with winding number $w = 0$ and $1$, as the final accuracy is more than 90%. Second, the trend of the loss function and the tripartite information is similar to that in (staggered) magnetization learning: At the early stage of the training, the loss decreases rapidly and the tripartite information increases. In the later stage, the tripartite information decreases again. The trend is robust when different initializations are averaged. However, we note the tripartite information is slightly more volatile in the later stage than that in the (staggered) magnetization learning, which is

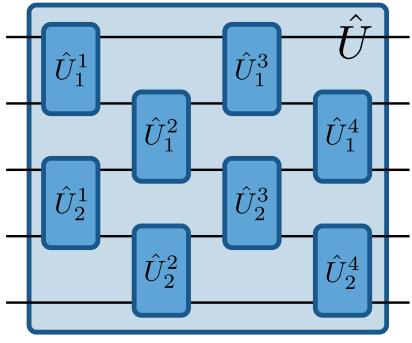

FIG. 7. Schematic of a quantum circuit with brick-wall geometry. Here the network has $n = 5$ qubits and depth $l = 4$. All these two-qubit gates form a giant unitary transformation $\hat{U}$. The $i$-th two-qubit gate in the $d$-th layer is denoted as $\hat{U}_i^d$

reflected by a second local maximum of the averaged $I_3$ around 350 epochs in Fig. 6. Because this behavior does not appear in other tasks, we believe it is not as universal and leave the in-depth understanding of this behavior for future research.

Compared with the (staggered) magnetization task, the input wavefunction here is a product state and is essentially classical, and the target is now a binary label instead of a real number. Despite the very different nature of this task, the empirical correlation between the NN performance and the tripartite information still holds. This suggests the generality of the two-stage training dynamics of quantum NNs.

## IV. GRADIENTS IN QUANTUM NNS

In this section, we report the method of computing gradients of quantum NNs in this work.

### A. In Classical Simulations

A schematic of the quantum NN with $n = 5$ qubits and depth $l = 4$ is shown in Fig. 7. The $i$-th two-qubit gate in the $d$-th layer is denoted as $\hat{U}_i^d$. Assuming $n$ is odd, here $i = 1, 2 \ldots (n-1)/2$. It follows the giant unitary $\hat{U}$ is the composition of $\hat{U}_i^d$:

$$\hat{U} = \left( \prod_{i=1}^{(n-1)/2} \hat{U}_i^l \right) \ldots \left( \prod_{i=1}^{(n-1)/2} \hat{U}_i^2 \right) \left( \prod_{i=1}^{(n-1)/2} \hat{U}_i^1 \right) \equiv \prod_{d=1}^{l} \left( \prod_{i=1}^{(n-1)/2} \hat{U}_i^d \right). \tag{16}$$

The order of unitaries within a layer does not matter because these unitaries are applied on non-overlapping qubits.

In general, each two-qubit gate $\hat{U}_i^d$ is a $4 \times 4$ matrix in the SU(4) group and can be parametrized by 15 parameters. However, as explained in the main text, in this work we restrict $\hat{U}_i^d$ to SO(4) with 6 Euler angles: Generally, a matrix in SO(4) can be parametrized by a vector $\boldsymbol{\theta}$ with 6 components [3]:

$$\hat{U}_{\mathrm{SO(4)}} = O_{34}(\theta_1) O_{23}(\theta_2) O_{12}(\theta_3) O_{34}(\theta_4) O_{23}(\theta_5) O_{34}(\theta_6). \tag{17}$$

Here $O_{ij}(\theta) \equiv \exp(\theta J_{ij})$ is a rotation in the $ij$ plane: $J_{ij}$ an antisymmetric matrix with $ij$ ($ji$) element equal to 1 ($-1$) and all other elements zero. As a result there are $l(n-1)/2$ independent vectors $\boldsymbol{\theta}_i^d$ and thus $6l(n-1)/2$ independent parameters in total to fully describe the quantum NN.

To be concrete, in the following, we use magnetization learning as the example. The staggered magnetization learning and winding number learning are similar. The loss function in magnetization learning is

$$\mathcal{L} = \frac{1}{N} \sum_{\alpha=1}^{N} \left| \langle G^\alpha | \hat{U}^\dagger \sigma_{(n+1)/2}^x \hat{U} | G^\alpha \rangle - M_z^\alpha \right|. \tag{18}$$

The gradient of $\mathcal{L}$ with respect to $\theta_{j,a}^d$, $a = 1, \ldots, 6$ is

$$\frac{\partial \mathcal{L}}{\partial \theta_{j,a}^d} = \frac{1}{N} \sum_{\alpha=1}^{N} \mathrm{sgn} \left( \langle G^\alpha | \hat{U}^\dagger \sigma_{(n+1)/2}^x \hat{U} | G^\alpha \rangle - M_z^\alpha \right) \frac{\partial \langle G^\alpha | \hat{U}^\dagger \sigma_{(n+1)/2}^x \hat{U} | G^\alpha \rangle}{\partial \theta_{j,a}^d}. \tag{19}$$

The gradient of the network output can be further simplified as

$$\frac{\partial}{\partial \theta_{j,a}^d} \langle G^\alpha | \hat{U}^\dagger \sigma_{(n+1)/2}^x \hat{U} | G^\alpha \rangle$$

$$= \langle G^\alpha | \hat{U}^\dagger \sigma_{(n+1)/2}^x \frac{\partial \hat{U}}{\partial \theta_{j,a}^d} | G^\alpha \rangle + \text{h.c.}$$

$$= \langle G^\alpha | \hat{U}^\dagger \sigma_{(n+1)/2}^x \frac{\partial}{\partial \theta_{j,a}^d} \left[ \prod_{d'=1}^{l} \left( \prod_{i=1}^{(n-1)/2} \hat{U}_i^{d'} \right) \right] | G^\alpha \rangle + \text{h.c.}$$

$$= \langle G^\alpha | \hat{U}^\dagger \sigma_{(n+1)/2}^x \left( \prod_{i=1}^{(n-1)/2} \hat{U}_i^l \right) \dots \left( \hat{U}_1^d \hat{U}_2^d \dots \frac{\partial \hat{U}_j^d}{\partial \theta_{j,a}^d} \dots \hat{U}_{\frac{n-1}{2}}^d \right) \dots \left( \prod_{i=1}^{(n-1)/2} \hat{U}_i^1 \right) | G^\alpha \rangle + \text{h.c.}, \tag{20}$$

where, $\partial \hat{U}_j^d / \partial \theta_{j,a}^d$ can be further simplified using Eq. (17). For example,

$$\frac{\partial \hat{U}_j^d}{\partial \theta_{j,4}^d} = O_{34}(\theta_{j,1}^d) O_{23}(\theta_{j,2}^d) O_{12}(\theta_{j,3}^d) J_{34} O_{34}(\theta_{j,4}^d) O_{23}(\theta_{j,5}^d) O_{34}(\theta_{j,6}^d). \tag{21}$$

Gradients with respect to other components $a$ can be computed in the similar way by adding an additional corresponding $J$ matrices.

In this work, we directly compute the gradient according to Eqs. (19), (20) and (21) in the classical simulation.

## B. In Real Quantum NNs

In a real quantum NN, this gradient could instead be determined through the measurement of the following Hermitian operator:

$$\hat{g}_{j,a}^d = \sigma_{(n+1)/2}^x \frac{\partial \hat{U}}{\partial \theta_{j,a}^d} \hat{U}^\dagger + \text{h.c.}. \tag{22}$$

It is straightforward to see that

$$\langle G^\alpha | \hat{U}^\dagger \hat{g}_{j,a}^d U | G^\alpha \rangle = \frac{\partial}{\partial \theta_{j,a}^d} \langle G^\alpha | \hat{U}^\dagger \sigma_{(n+1)/2}^x \hat{U} | G^\alpha \rangle.$$

However, this operator is generally non-local and is hard to measure.

Alternatively, one could perform the following three measurements [4, 5]:

1. Measure the output of the quantum NN normally with the original parameter $\theta_i^d$. The result is denoted as $o_1 \equiv \langle G^\alpha | \hat{U}^\dagger \sigma_{(n+1)/2}^x \hat{U} | G^\alpha \rangle$;

2. Measure the output of the quantum NN with $\theta_{j,a}^d$ replaced by $\theta_{j,a}^d + \pi/4$. The result is denoted as $o_2$;

3. Measure the output of the quantum NN with $\theta_{j,a}^d$ replaced by $\theta_{j,a}^d + \pi/2$. The result is denoted as $o_3$.

It follows the desired gradient is

$$\frac{\partial}{\partial \theta_{j,a}^d} \langle G^\alpha | \hat{U}^\dagger \sigma_{(n+1)/2}^x \hat{U} | G^\alpha \rangle = 2o_2 - o_1 - o_3. \tag{23}$$

The reason is that if we focus on some specific $\theta_{j,a}^d$, we have

$$o_1 = \left\langle \dots O_{p,p+1}^\dagger(\theta_{j,a}^d) \dots O_{p,p+1}(\theta_{j,a}^d) \dots \right\rangle, \tag{24}$$

$$o_2 = \left\langle \dots O_{p,p+1}^\dagger(\theta_{j,a}^d + \pi/4) \dots O_{p,p+1}(\theta_{j,a}^d + \pi/4) \dots \right\rangle$$

$$= \left\langle \dots \left[ (1 + J_{p,p+1}) O_{p,p+1}(\theta_{j,a}^d) \right]^\dagger \dots (1 + J_{p,p+1}) O_{p,p+1}(\theta_{j,a}^d) \dots \right\rangle / 2, \tag{25}$$

$$o_3 = \left\langle \dots O_{p,p+1}^\dagger(\theta_{j,a}^d + \pi/2) \dots O_{p,p+1}(\theta_{j,a}^d + \pi/2) \dots \right\rangle$$

$$= \left\langle \dots \left[ J_{p,p+1} O_{p,p+1}(\theta_{j,a}^d) \right]^\dagger \dots J_{p,p+1} O_{p,p+1}(\theta_{j,a}^d) \dots \right\rangle. \tag{26}$$

Here $p(p+1)$ is the rotation plane associated with $a$. As a result:

$$2o_2 - o_1 - o_3 = \left\langle \ldots O_{p,p+1}^\dagger(\theta_{j,a}^d) \ldots J_{p,p+1} O_{p,p+1}(\theta_{j,a}^d) \ldots \right\rangle + \text{h.c.} = \frac{\partial}{\partial \theta_{j,a}^d} \left\langle G^\alpha \right| \hat{U}^\dagger \sigma_{(n+1)/2}^x \hat{U} \left| G^\alpha \right\rangle . \quad (27)$$

The above method can be easily generalized to $\mathrm{SU}(4)$ as well.

---



\end{document}